\documentclass[12pt]{article}
 \usepackage{graphicx}
 \usepackage[cp1251]{inputenc} %% Windows
 \usepackage[russian]{babel}   %%

\begin{document}
 \noindent {\footnotesize\it Astronomy Letters, 2016, Vol. 42, No. 1, pp. 1--9.}
 \newcommand{\dif}{\textrm{d}}

 \noindent
 \begin{tabular}{llllllllllllllllllllllllllllllllllllllllllllll}
 & & & & & & & & & & & & & & & & & & & & & & & & & & & & & & & & & & & & & \\\hline\hline
 \end{tabular}

  \vskip 1.0cm
  \centerline{\bf Analysis of the Z Distribution of Young Objects in the Galactic Thin Disk}
  \bigskip
  \centerline{V.V. Bobylev and A.T. Bajkova}
  \bigskip
  \centerline{\small\it Pulkovo Astronomical Observatory, St. Petersburg,  Russia}
  \bigskip
  \bigskip
{\bf Abstract}—We have obtained new estimates of the Sun's
distance from the symmetry plane $Z_\odot$ and the vertical disk
scale height $h$ using currently available data on stellar OB
associations, Wolf-Rayet stars, HII regions, and Cepheids. Based
on individual determinations, we have calculated the mean
$Z_\odot=-16\pm2$~pc. Based on the model of a self-gravitating
isothermal disk for the density distribution, we have found the
following vertical disk scale heights: $h = 40.2\pm2.1$~pc from OB
associations, $h = 47.8\pm3.9$ pc from Wolf-Rayet stars,
$h=48.4\pm2.5$~pc from HII regions, and $h = 66.2\pm1.6$~pc from
Cepheids. We have estimated the surface, $\sum=6$~kpc$^{-2}$, and
volume, $D(Z_\odot) = 50.6$~kpc$^{-3}$, densities from a sample of
OB associations. We have found that there could be $\approx$5000
OB associations in the Galaxy.

%DOI: 10.1134/S1063773716010023

\section*{INTRODUCTION}
The Sun is known to be located not exactly in the Galactic plane
but to the north of it at a distance of about 20 pc. Two terms are
used: (a) the Sun’s height above the Galactic plane and (b) the
Sun’s position relative to the symmetry plane. The height is
positive and is usually designated as $h_\odot$. The Sun’s
position relative to the symmetry plane differs from the height
only by its sign and is usually designated as $Z_\odot$. In this
paper, we prefer to use $Z_\odot$. There are a number of
stellar-astronomy problems whose solution requires knowing an
accurate value of this quantity. For example, these include the
problems of constructing the Galactic orbit of the Sun or
establishing a highly accurate system of Galactic coordinates.

%%%%%%%%%%%%%%
 \begin{table}[t]         % t~1.
 \caption[]{\small
 Determinations of $Z_\odot$ by different authors from different data}
  \begin{center}\label{t:01}   \small %\footnotesize\baselineskip=0.1ex
  \begin{tabular}{|r|c|l|c|}\hline
                   Reference & $Z_\odot$, pc & Sample  \\\hline
               Toller (1990) & $-12.8\pm2.9$ & properties of dust from Pioneer 10 data \\
      Conti and Vacca (1990) & $-15\pm3$     & 101 Wolf-Rayet stars, $r<4.5$~kpc \\
      Brand and Blitz (1993) & $-13\pm7$     & 64 molecular clouds, $r:$ 0.7--2~kpc \\
 Humphreys and Larsen (1995) & $-20.5\pm3.5$ & optical star counts \\
    Hammersley et al. (1995) & $-15.5\pm3~~$ & IR COBE, IRAS, and TMGS data \\
        Binney et al. (1997) & $-14\pm4$     & IR COBE/DIRBE data  \\
 M\'endez, van Altena (1998) & $-27\pm3$     & star counts at low latitudes \\
     Maiz-Apell\'aniz (2001) & $-24.2\pm1.7$ & 3382 Hipparcos OB stars, $r<350$~pc \\
                 Reed (2006) & $-19.6\pm2.1$ & 2488 OB stars, $r<1.2$~kpc  \\
      Piskunov et al. (2006) & $-22\pm4$     & 254 open clusters, $r<850$~pc \\
                Yoshi (2007) & $-17\pm3$     & 537 open clusters, $r<4$~kpc \\
              Bobylev (2013) & $-23\pm5$     & 365 classical Cepheids, $r<4$~kpc \\\hline
                        Mean & $-18.6\pm1.4$ & \\\hline
  \end{tabular}\end{center}\end{table}
%%%%%%%%%%%%%%%%%%%%%%%%%%%%%%%%%%%%%%%%%%%%%%%%%%%%%%%%%%%%%%%%%

As a rule, the vertical disk scale height $h$ is determined
simultaneously with the constant $Z_\odot$. When analyzing the
distribution of young objects, we are dealing with the properties
of the Galactic thin disk. Different authors use different models
of the density distribution in the disk. This is either the model
of an exponential density distribution (see, e.g., Joshi 2007), or
the model of a self-gravitating isothermal disk (see, e.g., Conti
and Vacca 1990), or the Gaussian model (see, e.g.,
Maiz-Apell\'aniz 2001). Other models are also applied (Rosslowe
and Crowther 2015). Ultimately, such determinations are important
for constructing a present-day model of the Galactic disk.

Different authors have repeatedly determined $Z_\odot$ using young
O- and B-type stars, open clusters, Cepheids, infrared sources,
molecular clouds, and other objects. The methods of estimating
$Z_\odot$ are also very diverse. One of the first reviews of such
determinations was published by van Tulder (1942); he found
$Z_\odot= -13.5\pm1.9$ pc as a mean value from several samples of
various stars.

A review of present-day determinations can be found, for example,
in Humphreys and Larsen (1995) or Reed (2006). Selected results
are given in Table 1, which we will comment on in more detail.

Toller (1990) studied the distribution of interstellar Galactic
dust on the celestial sphere based on Pioneer 10 spacecraft
observations. He calculated the value of $Z_\odot$ in Table 1 as
the mean of three results obtained independently.

Conti and Vacca (1990) estimated the photometric distances of 157
Wolf–Rayet stars. These were all the known stars of this class at
that time. To determine $Z_\odot$, they used 101 stars from the
solar neighborhood with a radius $r<4.5$~kpc.

Brand and Blitz (1993) found $Z_\odot$ using data on the
distribution of 64 molecular clouds in the range of heliocentric
distances 0.7--2 kpc. They did not consider the nearest clouds,
closer than 0.7 kpc, to eliminate the influence of the Gould Belt.
The distances to the clouds were estimated photometrically.

Hammersley et al. (1995) used infrared data from the COBE (COsmic
Background Explorer) and IRAS (InfraRed Astronomical Satellite,
1988) satellites and the TMGS (Two-Micron Galactic Survey;
Garz\'on et al. 1993) catalog for their star counts.

Humphreys and Larsen (1995) made the star counts in 12 Palomar Sky
Survey fields near the North and South Galactic poles. Stars of
the thin Galactic disk were assumed to form the basis of the
sample. The ratio of the numbers of stars in different hemispheres
was derived directly from the counts. The $Z_\odot$ estimate
followed from a comparison with the model distribution of stars in
the Galactic disk. The model of the Galaxy by Bahcall and Soneira
(1980) was used for this purpose.

In contrast, M\'endez and van Altena (1998) made the star counts
at low Galactic latitudes. The Guide Star Catalog (Lasker et al.
1990) served as the basis. In this zone, the results of comparing
the star counts with the model distribution of stars depend
strongly on the proper allowance for interstellar extinction.
Therefore, M\'endez and van Altena (1988) first developed their
model of the 3D distribution of absorbing matter.

Binney et al. (1997) used infrared photometric data from the DIRBE
space experiment performed on the COBE satellite. They mapped the
surface brightness distribution in the inner Galaxy
($|l|\leq30^\circ, |b|\leq15^\circ$) by taking into account the
model of a central bar. By analyzing the observed distribution,
they could reach a conclusion about $Z_\odot$.

Maiz-Apell\'aniz (2001) used 3382 stars of spectral types from O
to B5 with their trigonometric parallaxes from the Hipparcos
(1997) catalogue. These stars are located in the solar
neighborhood with a radius $r<350$~pc and were selected under the
condition $|b|>5^\circ;$ therefore, the stars lying directly in
the Galactic plane, where the interstellar extinction is great,
were discarded. The parallaxes were corrected for the Lutz-Kelker
bias. The observed distribution of stars was fitted by the model
of a self-gravitating isothermal disk with two parameters:
$Z_\odot$ and the vertical disk scale height $h.$

Reed (2006) estimated $Z_\odot$ using 2488 OB stars from the solar
neighborhood with a radius $r<1.2$ kpc. The distances to these
stars were determined photometrically (with a relative error of
about 20\%).

Piskunov et al. (2006) estimated $Z_\odot$ based on a sample of
open stars clusters from the ASCC-2.5 catalog in the solar
neighborhood with a radius of 850 pc. They showed the catalog of
program clusters to be complete at this distance.

Joshi (2007) analyzed the spatial distribution of 537 fairly young
(younger than 300 Myr) open star clusters from the solar
neighborhood with a radius $r<4$~kpc. The distance and age
estimates were taken from various catalogs. The clusters
associated with the Gould Belt were excluded from consideration.

Bobylev (2013) considered the spatial distribution of classical
Cepheids in the solar neighborhood with a radius $r<4$~kpc. The
distances to the Cepheids were estimated based on the well-known
period--luminosity relation. The relative error in the stellar
distances determined by this method is 10\%--15\%. No constraints
were imposed on the pulsation period; the mean age of the Cepheids
in this sample can be roughly estimated to be 100 Myr.

It can be seen from Table 1 that the model--dependent estimates
(star counts, columns 4--7 in the Table) and the results of direct
calculations agree satisfactorily between themselves. There is no
noticeable correlation between $Z_\odot$ and the age of objects.
The last column of Table 1 gives the mean $Z_\odot$ with the error
of the mean calculated from 12 results.

The goal of this paper is to redetermine $Z_\odot$ and $h$ from
the currently available data on young objects. These are stellar
OB associations, HII regions, Wolf-Rayet stars, and Cepheids.
Galactic masers with measured trigonometric parallaxes located in
the regions of active star formation also belong to them.

\section*{METHODS}
In the case of an exponential density distribution, the observed
histogram of the distribution of objects along the $Z$ coordinate
axis is described by the expression
 \begin{equation}
  N(Z)=N_1 \exp \biggl(-{|Z-Z_\odot|\over h_1} \biggr),
 \label{elliptic}
 \end{equation}
where $N_1$ is the normalization coefficient.

If the model of a self-gravitating isothermal disk is used for the
density distribution, then the observed frequency distribution of
objects along the $Z$ axis is described by the formula (Spitzer
1942; Conti and Vacca 1990)
 \begin{equation}
  N(Z)=N_2{\hbox { sech}}^2 \biggl({Z-Z_\odot\over \sqrt2~h_2}\biggr).
 \label{self-grav}
 \end{equation}
When the results are compared, it should be kept in mind that in
model (2) different authors use either two,
 $N(Z)=N_2{\hbox { sech}}^2 [(Z-Z_\odot)/2h]$
 (Maiz-Apell\'aniz,~2001; Buckner and Froebrich,~2014), or
one,
 $N(Z)=N_2{\hbox { sech}}^2 [(Z-Z_\odot)/h]$
 (Marshall et al., 2006), as the coefficient in the denominator.

Finally, the observed frequency distribution of objects along the
$Z$ axis for the Gaussian model is described by the formula
 \begin{equation}
  N(Z)=N_3\exp\biggl[-{1\over 2}\biggl({Z-Z_\odot\over h_3}\biggr)^2\biggr].
 \label{Gauss}
 \end{equation}

Interestingly, when analyzing the $Z$ distribution of OB5 stars
based on Eqs. (2) and (3), Maiz-Apell\'aniz (2001) found the
vertical scale height $h_i$ to be determined with smaller errors
based on Eq. (2). In contrast, when studying OB stars based on
Eqs. (1) and (3), Elias et al. (2006) showed the vertical scale
height $h$ to be determined more accurately based on Eq. (3).
Therefore, the model of a self-gravitating isothermal disk (2) is
the most attractive one among the three described models.

The vertical disk scale heights $h_i$ determined by different
authors from different data based on models (1), (2), and (3) are
given in Table 2. Stothers and Frogel (1974) studied $\approx$1000
O--B5 stars closely associated with the Gould Belt. Reed (2000)
considered up to 4000 OB stars with photometric distances in the
solar neighborhood with a radius of 4 kpc. The analysis in Elias
et al. (2006) was based on a sample of 553 Hipparcos O--B6 stars,
with the Gould Belt stars having been excluded. The sample in
Joshi (2007) consisted of 2030 OB stars and contained no Gould
Belt stars.

The collected results lead us to conclude that the vertical disk
scale heights determined using different models (a) agree
satisfactorily between themselves if they are applied to samples
of the same age and (b) differ noticeably in the case of samples
of different ages.

%%%%%%%%%%%%%%
 \begin{table}[t]            % t~2.
 \caption[]{\small
 Determinations of the vertical disk scale height $h_i, i=1,2,3$
 from data based on three models: (1), (2), and (3)}
  \begin{center}
  \label{t:02}   \small %\footnotesize\baselineskip=0.1ex
  \begin{tabular}{|r|c|c|c|l|c|}\hline
               Reference & $h_1,$~pc  & $h_2,$~pc    & $h_3,$~pc    & Sample   \\\hline
 Stothers and Frogel (1974) & $ 46\pm7$  & ---          & $ 48\pm2$    & B0-B5, 0--200~pc \\
 Stothers and Frogel (1974) & $ 70\pm3$  & ---          & $ 61\pm3$    & B0-B5, 0--800~pc \\
   Conti and Vacca (1990) & ---        & $45\pm5$     & ---          & WR stars \\
              Reed (2000) & $45$       & ---          & ---          & OB stars \\
  Maiz-Apell\'aniz (2001) & ---        & $34.2\pm2.5$ & $62.8\pm4.7$ & OB stars (*) \\
    Bonatto et al. (2006) & $ 48\pm3$  & ---          & ---          & SCs, $<$200~Myr \\
    Bonatto et al. (2006) & $150\pm27$ & ---          & ---          & SCs, 0.2--1~Gyr \\
      Elias et al. (2006) & $34\pm2$   & ---          & $30\pm2$     & OB stars \\
   Piskunov et al. (2006) & $56\pm3$   & ---          & $74\pm3$     & 254 OSCs \\
             Yoshi (2007) & $56.9^{+3.8}_{-3.4}$ & --- & ---         & 537 OSCs \\
             Yoshi (2007) & $61.4^{+2.7}_{-2.4}$ & --- & ---         & OB stars \\
 \hline
 \end{tabular}
 \end{center}
 {\small %\footnotesize\baselineskip=0.1ex
 WR: Wolf-Rayet stars; OSCs: open star clusters;
 (*): the value of $h_2$ from Maiz-Apell\'aniz (2001) should be multiplied by
 $\sqrt 2$ for its comparison with the result from Conti and Vacca
(1990).
 }
 \end{table}
%%%%%%%%%%%%%%%%%%%%%%%%%%%%%%%%%%%%%%%%%%%%%%%%%%%%%%%%%%%%%%%%%

\section*{DATA}
(1) We use the data on known OB associations with reliable
distance estimates from Mel\'nik and Dambis (2009). The distances
derived by these authors were reconciled with the Cepheid scale.
The catalog contains 91 OB associations; the distances to them do
not exceed $r = 3.5$~kpc. The designations of associations
including stars of spectral types up to B9 (for example, Cyg OB9)
are encountered in the catalog. On this basis, we can conclude
that the upper age limit is $\approx$60--80 Myr. Mel\'nik and
Dambis (2009) estimated the errors in the distances to these OB
associations to be 10--15\%.

(2) The Wolf-Rayet stars described by Rosslowe and Crowther (2015)
constitute another sample. This is the most complete sample of
Galactic stars of this type to date. Note that Rosslowe and
Crowther studied their $Z$ distribution. A model based on the
Cauchy distribution was used to determine the disk scale height.
Rosslowe and Crowther (2015) themselves estimated new distances
for 246 stars. They added 108 more stars with known distances from
the van der Hucht (2001) catalog to them and produced a sample of
354 stars. All these stars are located within 35 kpc of the Sun. A
new analysis of these data using various constraints is of
interest. First of all, very distant stars need to be excluded,
because the effect due to a large-scale warp of the Galactic thin
disk, in particular, a warp of the neutral hydrogen layer, must
have a noticeable influence at great distances from the Sun (more
than 8--10 kpc). Note that Rosslowe and Crowther (2015) give
individual estimates of the random errors in the distances for 246
stars from their list. For the 108 stars from the van der Hucht
(2001) catalog that have no such estimates, we assume the distance
error to be 20\%.

(3) We use the sample of classical Cepheids belonging to our
Galaxy described by Mel'nik et al. (2015). This is the most
complete sample of such objects with known proper motions and
radial velocities to date. This sample contains 674 stars whose
distances were determined from the most recent calibrations using
both optical and infrared photometric observational data. We
assume the error in the distance to a star determined by the Monte
Carlo method to be 10\%.

(4)We use a sample of Galactic masers with measured trigonometric
parallaxes. These sources, associated with very young stars, are
located in regions of active star formation. Highly accurate
astrometric VLBI measurements of the trigonometric parallaxes and
proper motions have already been performed for more than 120 such
masers by several teams in the USA, Japan, Europe, and Australia
(Reid et al. 2014). The error in the stellar parallax determined
by this method is, on average, less than 10\%.

We compiled our sample based on a number of publications. The
paper of Xu et al. (2013) is devoted to masers in the Local Arm.
The review of Reid et al. (2014) contains data of 103 masers.
Subsequently, the publications of these authors with improvements
and additions devoted to the analysis of masers in individual
Galactic spiral arms appeared. The papers by Sanna et al. (2014),
Sato et al. (2014), Wu et al. (2014), Choi et al. (2014), and
Hachisuka et al. (2015) are devoted to the inner Galaxy, the
Scutum Arm, the Carina–Sagittarius Arm, the Perseus Arm, and the
Outer Arm, respectively. Having added the most recent published
results of astrometric measurements, we produced a sample
containing data on 130 masers.

(5) We supplemented the well-known catalog of HII regions (Russeil
2003) with new photometric estimates of the distances to several
star-forming regions collected in Russeil et al. (2007) and
Mois\'es et al. (2011). From this catalog we took only those
distances that were obtained photometrically. We assume the
relative error of such distances to be 20\%.

%%%%%%%%%%%%%%%%%%%%%%%%%%%%%%%%%%%%%%%%%%%%%%%%%%%%%%%%%% Fig. 1
 \begin{figure} {\begin{center}
 \includegraphics[width=136mm]{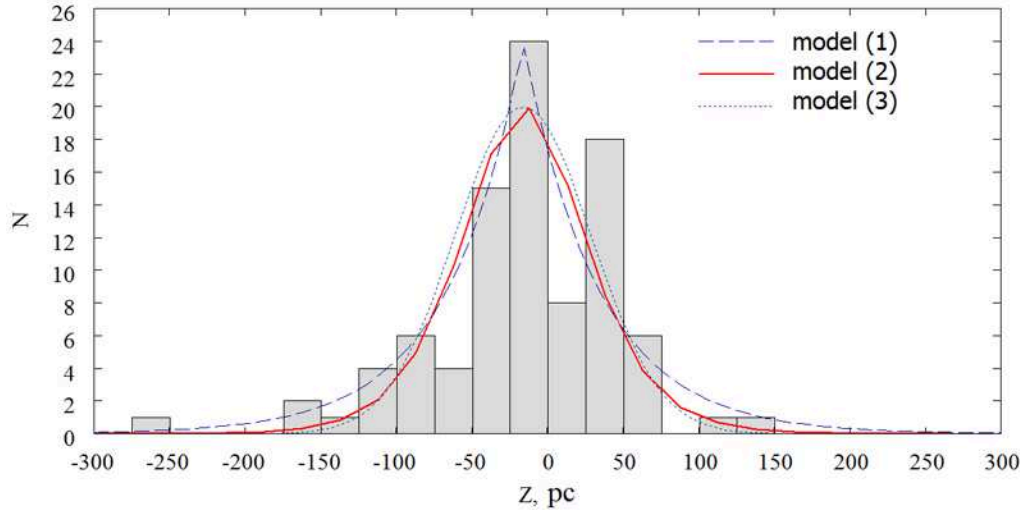}
 \caption{Histogram of the $Z$ distribution of OB associations. }
 \label{f1}
 \end{center} } \end{figure}
%%%%%%%%%%%%%%%%%%%%%%%%%%%%%%%%%%%%%%%%%%%%%%%%%%%%%%%%%%%%%%%%%

%%%%%%%%%%%%%%%%%%%%%%%%%%%%%%%%%%%%%%%%%%%%%%%%%%%%%%%%%%%%%%%%% Fig 2
 \begin{figure} {\begin{center}
 \includegraphics[width=165mm]{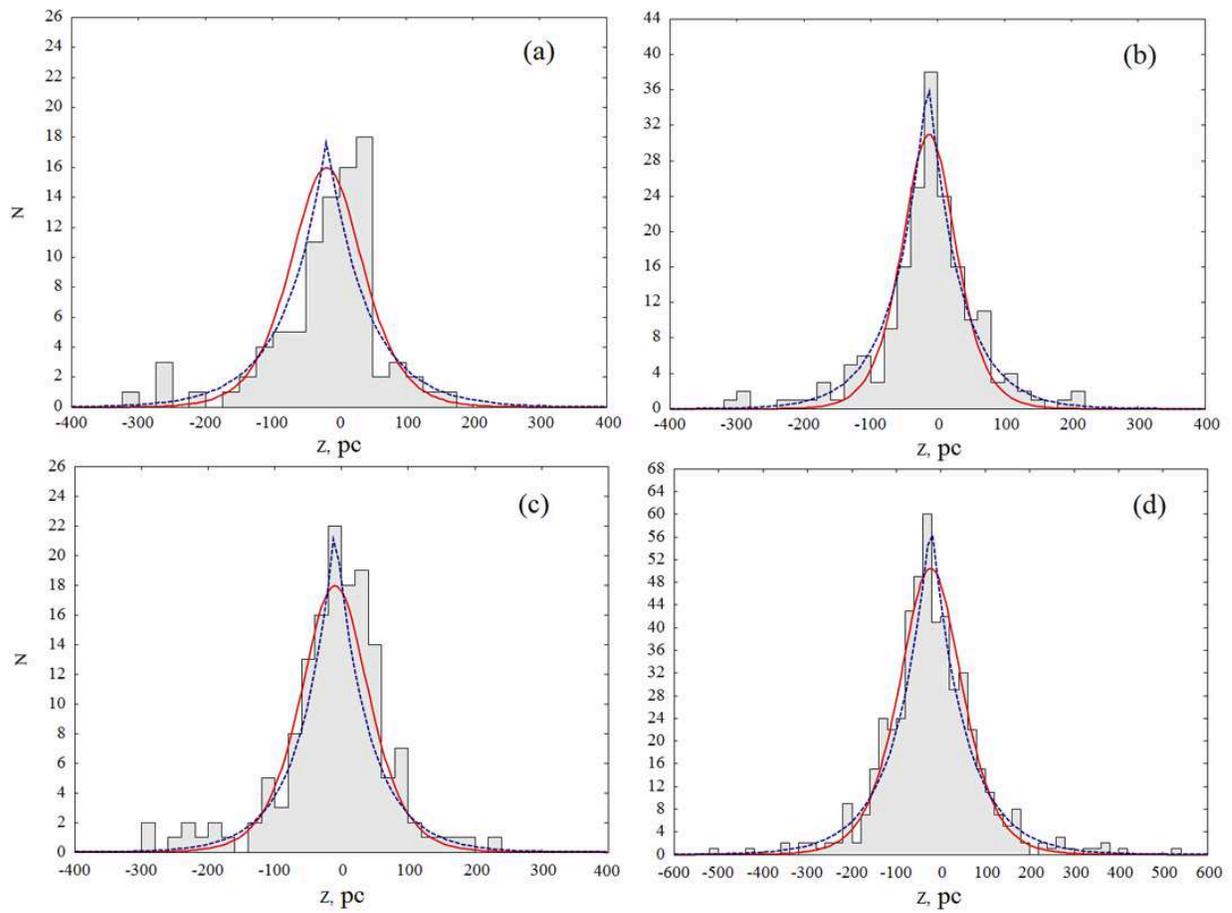}
 \caption{Histogram of the Z distribution of masers a), HII regions b),
Wolf-Rayet stars c), and Cepheids d); the dashed and solid lines
curves on all panels represent models (1) and (2), respectively. }
 \label{f2} \end{center} } \end{figure}
%%%%%%%%%%%%%%%%%%%%%%%%%%%%%%%%%%%%%%%%%%%%%%%%%%%%%%%%%%%%%%%%%

\section*{RESULTS AND DISCUSSION}
Table 3 presents the results of our determinations of the Sun’s
distance from the symmetry plane $Z_\odot$ and the vertical disk
scale height $h_i, i=1,2,3$ that we obtained based on models
(1)--(3) using the samples of various objects. These parameters
and their errors were found by fitting the models to the
histograms and through Carlo simulations. For this purpose, we
constructed the histograms with a step of 10 pc in $Z$ coordinate.
Figure 1 shows the histogram of the $Z$ distribution of OB
associations with a step of 25 pc for better clarity.

Note that we actually have two sets of data of different ages.
Masers, OB associations, HII regions, and Wolf–Rayet stars
constitute the group of the youngest objects, while Cepheids form
the second group of older objects. As can be seen from Table 3,
the values of $h_i$ found for the first group of objects agree
satisfactorily between themselves, except for the masers. We will
discuss the masers below. There is good agreement in the
determinations of $Z_\odot$ from the samples of all objects
without exceptions. The histograms of the $Z$ distribution of
masers, HII regions, Wolf–Rayet stars, and a sample of 496
Cepheids are presented in Fig. 2.

\subsection*{OB Associations}
The three model fitting curves constructed with the parameters
from Table 3 are plotted in Fig. 1. From the figure we can see no
significant differences between all three curves, especially
between curves (2) and (3). Different authors simply use different
models based on the formulated problems and personal preferences.
For example, the convenience of using model (1) lies in the fact
that the distribution of objects in a logarithmic scale is fitted
by two line segments (see, e.g., Fig. 3 in Piskunov et al.
(2006)). Since curves (2) and (3) virtually coincide, we
restricted ourselves to the two models (1) and (2) in Fig. 2.

\subsection*{HII Regions}
The results of our analysis of HII regions are valuable in that we
use only the distances derived photometrically from their central
stars. It is interesting to compare our results to those obtained
using indirect distance estimation methods.

Paladini et al. (2004) analyzed the $Z$ distribution of 550 HII
regions. Their distances were estimated partly by the kinematic
method from the Galactic rotation curve and partly from the
luminosity–diameter correlation. Therefore, their quality is not
high. Nevertheless, these authors found the negative mean
coordinate $Z_\odot=-11.3$~pc and the vertical disk scale height
$h_3=52$~pc that agree satisfactorily with those we found.

Bronfman et al. (2000) studied 748 star-forming regions and $H_2$
clouds distributed over the entire Galaxy. They used the kinematic
distances. These authors found $Z_\odot\approx-5$~pc and
$h\approx40$~pc (here, $h$ is the half-width at half-height of the
Gaussian distribution) from the HII regions and
$Z_\odot\approx-6$~pc and $h\approx60$~pc from the molecular
hydrogen clouds. Since a very wide solar neighborhood was used
here, significant differences in the $Z_\odot$ determinations from
northern ($0^\circ<l<180^\circ$) and southern
($180^\circ<l<360^\circ$) sky objects were found. In spite of
this, the vertical scale heights are in satisfactory agreement
with our estimates.

\subsection*{Wolf–Rayet Stars}
Out of the total number of 354 Wolf-Rayet stars, there are 148
stars in the solar neighborhood with a radius $r<4.5$~kpc. This
sample is a factor of 1.5 larger than that analyzed by Conti and
Vacca (1990). The value of $h_2=47.8\pm3.9$~pc we found is in
excellent agreement with the result, $h_2=45\pm5$~pc, obtained by
Conti and Vacca (1990).

Note that in comparison with all of the remaining samples, the
values of hi that we found from the Wolf-Rayet stars have the
largest errors. This is because the individual distance errors for
246 stars from the list by Rosslowe and Crowther (2015)
occasionally reach $\approx$30\%.

%%%%%%%%%%%%%%
 \begin{table}[t]                                     % t~3.
 \caption[]{\small
Determinations of the Sun’s distance from the symmetry plane
$Z_\odot$ and the vertical disk scale height $h_{i, ~i=1,2,3}$ in
the paper based on models: (\ref{elliptic})--(\ref{Gauss})}
  \begin{center}  \label{t:03}
  \begin{tabular}{|c|c|c|c|l|c|}\hline
 $Z_\odot,$~pc & $h_1,$~pc & $h_2,$~pc & $h_3,$~pc & Sample  \\\hline
 $-19\pm4$  &   $61\pm4$   &   $51\pm3$   &   $58\pm3$   & ~~90 masers, $r<4$~kpc \\
 $-16\pm2$  & $44.9\pm2.5$ & $40.2\pm2.1$ & $46.9\pm2.3$ & ~~91 OB associations, $r<3.5$~kpc \\
 $-15\pm3$  & $48.6\pm2.5$ & $48.4\pm2.5$ & $48.6\pm2.5$ &  187 HII regions, $r<4.5$~kpc \\
 $-10\pm4$  & $51.3\pm3.7$ & $47.8\pm3.9$ & $54.8\pm3.5$ &  148 Wolf--Rayet stars, $r<4.5$~kpc \\
 $-23\pm2$  & $70.2\pm2.4$ & $60.1\pm1.9$ & $69.6\pm2.4$ &  246 Cepheids, ${\overline t}\approx~$75~Myr, $r<4$~kpc \\
 $-24\pm2$  & $83.8\pm2.4$ & $72.5\pm2.3$ & $83.2\pm2.8$ &  250 Cepheids, ${\overline t}\approx$138~Myr, $r<4$~kpc \\
 $-23\pm2$  & $76.4\pm1.8$ & $66.2\pm1.6$ & $76.6\pm1.9$ &  496 Cepheids, ${\overline t}\approx$107~Myr, $r<4$~kpc \\
 \hline
 \end{tabular}\end{center}
 \end{table}
%%%%%%%%%%%%%%%%%%%%%%%%%%%%%%%%%%%%%%%%%%%%%%%%%%%%%%%%%%%%%%%%%

\subsection*{Cepheids}
When analyzing the spatial distribution of Cepheids, Bobylev
(2013) showed that the disk warp at distances $r>4$~kpc affects
significantly the calculation of $Z_\odot$. Therefore, we used
this constraint on the Cepheid distances. In the range of
heliocentric distances $r<4$~kpc, the catalog of Mel’nik et al.
(2015) contains 496 Cepheids with various pulsation periods $P.$
We additionally divided the entire sample into two equal (in the
number of stars) parts depending on the pulsation period with the
boundary $P=5.5$~days.

Several calibrations proposed to estimate the individual ages of
Cepheids ($t$) from the pulsation period are known. We used the
calibration from Efremov (2003),
\begin{equation}
 \log t=8.5-0.65 \log P,
\label{AGE-EFREM}
 \end{equation}
derived by this author from Cepheids belonging to open clusters in
the Large Magellanic Cloud. As a result, we have two samples of
Cepheids with mean ages $t\approx75$ and $\approx138$~Myr, while
the mean age of the entire sample of 496 stars is
$\approx107$~Myr. The theoretical calibration from Bono et al.
(2005), $\log t = 8.31-0.67\log P$, for Cepheids with a mean
metallicity of 0.02 typical of Galactic stars is also known. The
experience of its application shows (Bobylev and Bajkova 2012)
that, in this case, the mean age of Cepheids turns out to be a
factor of 1.5 younger than that based on calibration~(4).

As can be seen from Table~3, owing to the large number of
Cepheids, the values of $Z_\odot$ and $h$ found from them are
highly accurate. It can be seen from Fig. 2 that the distribution
of Cepheids is among the best ones from the standpoint of its
symmetry. The distribution of HII regions is only slightly
inferior to it, while the distributions of Wolf-Rayet stars and
masers are most asymmetric.

We can also see that the vertical scale height h increases with
increasing age of the sample stars. On the whole, this is in
agreement with the results of the analysis of open star clusters
with different ages performed, for example, by Bonatto et al.
(2006) or Piskunov et al. (2006). It should be noted that the
estimates of these authors disagree noticeably for the oldest open
clusters. For example, Piskunov et al. (2006) found $h_1 =
61\pm6$~pc for the sample of open clusters with ages $\log t>8.6$
(older than $\approx400$~Myr), which is half the value of
$h_1=150\pm27$~pc obtained by Bonatto et al. (2006) from a sample
of clusters with ages in the range 0.2--1 Gyr. The relation
between the vertical disk scale height and the age of stars found
by Buckner and Froebrich (2014) from the data on open star
clusters is of interest in this connection. As can be seen from
Fig. 6 of these authors, $h$ is about 40 pc for clusters with ages
of $\approx1$~Myr, 50 pc for clusters with ages of
$\approx10$~Myr, reaches about 100 pc at an age of $\approx1$~Gyr,
and only then does it begin to increase sharply to 550 pc at an
age of $\approx3.5$~Gyr.

\subsection*{Masers}
We used the data on maser sources that are associated mainly with
massive protostars located in regions of active star formation.
Therefore, one might expect that the values of $h$ found from the
sample of masers will not exceed those found from the HII regions,
OB associations, and Wolf-Rayet stars. However, but this does not
hold.

The sample of masers is peculiar in that their observations aimed
at determining the trigonometric parallaxes are performed from the
Earth’s northern hemisphere. The parallax was determined using the
Australian radio interferometer only for one source from the
southern hemisphere. As a result, only the first and second
quadrants are well filled in Galactic coordinates, while the
fourth quadrant still remains almost empty (Fig. 2 in Bobylev and
Bajkova (2014)).

We can conclude that to realize the existing high potential of
these data, we should wait for the appearance of trigonometric
parallax determinations with interferometers in the Earth’s
southern hemisphere. Such observations are currently being
performed.

\subsection*{The Mean Value of $Z_\odot$}
Based on the $Z_\odot$ determinations from Table 3, we calculated
the mean value of $Z_\odot=-16\pm2$~pc (the dispersion here is
4~pc, while 2~pc is the error of the mean). For this purpose, we
used of four samples: (1) OB associations, (2) HII regions, (3)
Wolf-Rayet stars, and (4) 496 Cepheids. This is a new estimate. It
agrees satisfactorily with the results listed in Table~1, in
particular, with the mean value of $Z_\odot=-18.6\pm1.4$~pc.

 \subsection*{Estimating the Number of OB Associations in the Galaxy}
Based on the derived parameters, we can calculate the surface,
$\sum (Z)$, and volume, $D(Z)$, densities of objects. For this
purpose, let us rewrite Eq.~(1) in terms of the density $D(Z)$:
 \begin{equation}
  D(Z)=D(Z_\odot) \exp \biggl(-{|Z-Z_\odot|\over h_1} \biggr).
 \label{elliptic-DZ}
 \end{equation}
The surface and volume densities are related by the relation
(Piskunov et al. 2006)
 \begin{equation}
  \sum(Z)=2D(Z_\odot) h_1 \biggl[1-\exp \biggl(-{|Z-Z_\odot|\over h_1}
  \biggr)\biggr],
 \label{elliptic-SSZ}
 \end{equation}
where $\sum(Z)$ denotes the surface density in a layer of
thickness $2|Z-Z_\odot|$. We can estimate the total number of
objects in the Galaxy $N_{tot}$ using the relation (Piskunov et
al. 2006)
 \begin{equation}
 N_{tot}=2\pi\sum\int^{R^{lim}}_0 \exp\biggl(-{R-R_\odot\over L_d}\biggr)R dR,
 \label{Total-NNZ}
 \end{equation}
where $R$ is the distance from the star to the Galactic center,
$R^{lim}$ is the Galactocentric radius of the disk, $L_d$ is the
radial disk scale length, and $\sum=N_f/\pi d^2_{xy}$ is the
surface density in the solar neighborhood found by assuming the
sample to be complete, where $d_{xy}$ is the sample completeness
radius in the $XY$ plane and $N_f$ is the number of objects in
this sample. Just as in Piskunov et al (2006), we take
$L_d=3.5$~kpc and $R_\odot=8$~kpc, which correspond to the model
from Bahcall and Soneira (1980).

We use this approach to calculate the total number of OB
associations, because we can easily compare the result with the
estimates of the total number of open clusters in the Galaxy. Such
estimates can be found, for example, in Piskunov et al. (2006) or
Bonatto et al. (2006). Actually, the Scorpio–Centaurus OB
association nearest to the Sun is known to contain about ten open
clusters; the picture in the younger association in Orion is
similar. Therefore, the expected number of young clusters in an
association must be about ten.

According to the data in Table 3 for the sample of OB
associations, we have $h_1=44.9$~pc and $Z_\odot=-16$~pc. We then
estimated our sample of OB associations to be complete in the
solar neighborhood with the radius $d_{xy}=1.87$~pc. This
neighborhood contains $N_f = 67$ associations. With these
parameters, we find the surface and volume densities to be
$\sum=6$ kpc$^{-2}$ and $D(Z_\odot)=50.6$ kpc$^{-3}$,
respectively.

As a result, we found that $N_{tot}=4949$ OB associations could be
within the disk of radius $R_{lim}=15$ kpc. Here, we see agreement
with the result of Piskunov et al. (2006), where the total number
of open clusters in the Galaxy was found under the same conditions
to be $N_{tot} = 93000$. Indeed, these estimates show that, on
average, one OB association contains $93 000/4949\approx19$
clusters.

Bonatto et al. (2006) estimated the total number of clusters in
the Galaxy to be 1.8--3.7$\times10^5$. In this case, one OB
association must contain from 36 to 74 clusters. Here, the
agreement with our results is considerably poorer.

Since OB associations can contain only young clusters, we can
conclude that our estimates agree with the results of other
authors in order of magnitude.

Note that our estimate of $N_{tot}=4949$ may be considered (with
some reservations) as an upper limit on the number of OB
associations. The Sun is located near two spiral arms, where the
orbits of gas clouds come close together, and the frequency of
collisions between clouds increases; such stimulated star
formation leads to the birth of OB associations. Furthermore, the
Sun may be located near the bar-induced Lindblad outer resonance.
Therefore, there may be more gas here than at other radii.

\section*{CONCLUSIONS}
We obtained new estimates of the Sun's distance from the symmetry
plane $Z_\odot$ and the vertical disk scale height $h$ using
samples of various objects. For this purpose, we took the
currently available data on (1) 91 stellar OB associations, (2)
187 HII regions, (3) 148 Wolf-Rayet stars, (4) 90 maser sources in
regions of active star formation, and (5) 496 classical Cepheids.

To each of these samples, we applied three models of the density
distribution: the model of an exponential distribution, the model
of a self-gravitating isothermal disk, and the model of a Gaussian
density distribution. We showed that the derived vertical disk
scale heights $h$ depend on the age of sample objects when using
any of these three models.

It turned out that the values of $h$ found from the sample of
masers deviate noticeably from this dependence. We concluded that
the values of $h$ found from the sample of masers are not
reliable. This is because the sample available to date is so far
dominated by northern-sky masers (Galactic longitudes
$0^\circ<l<180^\circ$). The remaining four samples have a uniform
longitude distribution; obtained quite reliable estimates from
them.

Based on the individual determinations of $Z_\odot$ from four
different samples (without the sample of masers), we calculated
the mean value of $Z_\odot=-16\pm2$ pc. This is a new estimate,
but, at the same time, it is in good agreement with other known
estimates of this quantity.

We showed the vertical disk scale height to be determined with the
smallest errors based on the model of a self-gravitating
isothermal disk (model~(2)), using which we found
 $h=40.2\pm2.1$~pc from OB associations,
 $h=47.8\pm3.9$~pc from Wolf-Rayet stars,
 $h=48.4\pm2.5$~pc from HII regions, and
 $h=66.2\pm1.6$~pc from Cepheids.

We estimated the surface and volume densities from the sample of
OB associations to be $\sum = 6$~kpc$^{-2}$ and
 $D(Z_\odot)=50.6$ kpc$^{-3}$, respectively. Based on
these estimates, we showed that there could be about 5000 OB
associations in the Galaxy.

\subsection*{ACKNOWLEDGMENTS}
We are grateful to the referees for their helpful
remarks that contributed to an improvement of this paper. This
work was supported by the ``Transitional and Explosive Processes
in Astrophysics'' Program P--41 of the Presidium of Russian
Academy of Sciences.

 \bigskip{REFERENCES}
 \medskip
 {\small

 1. J.N. Bahcall and R.M. Soneira, Astrophys. J. Suppl. Ser. 44, 73 (1980).

 2. J. Binney, O. Gerhard, and D. Spergel, Mon. Not. R. Astron. Soc. 288, 365 (1997).

3. V.V. Bobylev and A.T. Bajkova, Astron. Lett. 38, 638 (2012).

4. V.V. Bobylev, Astron. Lett. 39, 753 (2013).

5. V.V. Bobylev and A.T. Bajkova, Mon. Not. R. Astron. Soc. 437,
1549 (2014).

6. C. Bonatto, L.O. Kerber, E. Bica, and B.X. Santiago, Astron.
Astrophys. 446, 121 (2006).

7. G. Bono, M. Marconi, S. Cassisi, F. Caputo, W. Gieren, and G.
Pietrzynski, Astrophys. J. 621, 966 (2005).

8. J. Brand and L. Blitz, Astron. Astrophys. 275, 67 (1993).

 9. L. Bronfman, S. Casassus, J. May, and L.-\AA. Nyman, Astron.
Astrophys. 358, 521 (2000).

10. A.S.M. Buckner, and D. Froebrich, Mon. Not. R. Astron. Soc.
444, 290 (2014).

11. Y.K. Choi, K. Hachisuka, M.J. Reid, Y. Xu, A. Brunthaler, K.M.
Menten, and T.M. Dame, Astrophys. J. 790, 99 (2014).

12. P.S. Conti and W.D. Vacca, Astron. J. 100, 431 (1990).

13. Yu.N. Efremov, Astron.Rep. 47, 1000 (2003).

14. F. Elias, J. Cabrero-Can'o, and E.J. Alfaro, Astron. J. 131,
2700 (2006).

15. F. Garz\'on, P.L. Hammersley, T. Mahoney, X. Calbet, M.J.
Selby, and I. Hepburn, Mon. Not. R. Astron. Soc. 264, 773 (1993).

16. K. Hachisuka, Y.K. Choi, M.J. Reid, A. Brunthaler, K.M.
Menten, A. Sanna, and T.M. Dame, Astrophys. J. 800, 2 (2015).

17. P.L. Hammersley, F. Garz\'on, T. Mahoney, and X. Calbet, Mon.
Not. R. Astron. Soc. 273, 206 (1995).

18. K.A. van der Hucht, New Astron. Rev. 45, 135 (2001).

19. R.M. Humphreys and J.A. Larsen, Astron. J. 110, 2183 (1995).

20. Y.C. Joshi, Mon. Not. R. Astron. Soc. 378, 768 (2007).

21. B.M. Lasker, C.R. Sturch, B.J. McLean, J.L. Russell, H.
Jenkner, and M.M. Shara, Astron. J. 99, 2019 (1990).

22. J. Maiz-Apell\'aniz, Astron. J. 121, 2737 (2001).

23. D.J. Marshall, A.C. Robin, C. Reyl\'e, M. Schultheis, and S.
Picaud, Astron. Astrophys. 453, 635 (2006).

24. A.M. Mel’nik and A.K. Dambis, Mon. Not. R. Astron. Soc. 400,
518 (2009).

25. A.M. Mel’nik, P. Rautiainen, L.N. Berdnikov, A.K. Dambis, and
A.S. Rastorguev, Astron. Nachr. 336, 70 (2015).

 26. R.A. M\'endez and W.F. van Altena, Astron. Astrophys. 330, 910 (1998).

27. A.P. Mois\'es, A. Damineli, E. Figueredo, R.D. Blum, P.S.
Conti, and C.L. Barbosa, Mon. Not. R. Astron. Soc. 411, 705
(2011).

28. R. Paladini, R.D. Davies, and G. DeZotti, Mon. Not. R. Astron.
Soc. 347, 237 (2004).

29. A.E. Piskunov, N.V. Kharchenko, S. Roeser, E. Schilbach, and
R.-D. Scholz, Astron. Astrophys. 445, 545 (2006).

30. B.C. Reed, Astron. J. 120, 314 (2000).

31. B.C. Reed, J.R. Astron. Soc. Canada 100, 146 (2006).

32. M.J. Reid, K.M. Menten, A. Brunthaler, X.W. Zheng, T.M. Dame,
Y. Xu, Y. Wu, B. Zhang, A. Sanna, M. Sato, et al., Astrophys. J.
783, 130 (2014).

33. C.K. Rosslowe and P.A. Crowther, Mon. Not. R. Astron. Soc.
447, 2322 (2015).

34. D. Russeil, Astron. Astrophys. 397, 133 (2003).

35. D. Russeil, C. Adami, and Y.N. Georgelin, Astron. Astrophys.
470, 161 (2007).

36. A. Sanna, M.J. Reid, K.M. Menten, T.M. Dame, B. Zhang, M.
Sato, A. Brunthaler, L. Moscadelli, and K. Immer, Astrophys. J.
781, 108 (2014).

37. M. Sato, Y.W. Wu, K. Immer, B. Zhang, A. Sanna, M.J. Reid,
T.M. Dame, A. Brunthaler, and K.M. Menten, Astrophys. J. 793, 72
(2014).

38. L. Spitzer, Astrophys. J. 95, 329 (1942).

39. R. Stothers and J.A. Frogel, Astron. J. 79, 456 (1974).

40. G.N. Toller, IAU Symp. 139, 21 (1990).

41. J.J.M. van Tulder, Bull. Astron. Inst. Netherlands 9, 315
(1942).

42. Y.W. Wu, M. Sato, M.J. Reid, L. Moscadelli, B. Zhang, Y. Xu,
A. Brunthaler, K.M. Menten , T.M. Dame, and X.W. Zheng, Astron.
Astrophys. 566, 17 (2014).

43. Y. Xu, J.J. Li, M.J. Reid, X.W. Zheng, A. Brunthaler, L.
Moscadelli, T.M. Dame, and B. Zhang, Astrophys. J. 769, 15 (2013).

44. IRAS Point Source Catalog (NASA, Washington, DC, 1988).

45. The Hipparcos and Tycho Catalogues, ESA SP--1200 (1997).

 }
\end{document}